\documentclass[12pt]{article}
\usepackage{amsmath,amssymb,graphicx,mathrsfs}
\usepackage{hyperref}
\newcommand{\be}{\begin{equation}}
\newcommand{\ee}{\end{equation}}
\newcommand{\bea}{\begin{eqnarray}}
\newcommand{\eea}{\end{eqnarray}}

\def\({\left(} \def\){\right)}

\def\revise#1       {\raisebox{-0em}{\rule{3pt}{1em}}
                     \marginpar{\raisebox{.5em}{\vrule width3pt\
                     \vrule width0pt height 0pt depth0.5em
                     \hbox to 0cm{\hspace{0cm}{%
                     \parbox[t]{4em}{\raggedright\footnotesize{#1}}}
\hss}}}}

\begin{document}

\title{The shear diffusion coefficient for generalized theories of
gravity}
\author{  Ram Brustein \\   Department of Physics, Ben-Gurion
University,\\
    Beer-Sheva, 84105 Israel \\ E-mail: ramyb@bgu.ac.il\\ \\
A.J.M. Medved \\ Physics Department, University of Seoul, \\
Seoul 130-743 Korea \\
    E-mail: allan@physics.uos.ac.kr}
\date{}
\maketitle


\abstract{
Near the horizon of a black brane in Anti-de Sitter (AdS) space and
near the AdS boundary, the long-wavelength fluctuations of
the metric exhibit hydrodynamic behaviour. The gauge-gravity duality
then relates the boundary hydrodynamics for generalized gravity to that of
gauge theories with large finite values of 't Hooft coupling.
We discuss, for this framework, the
hydrodynamics of the shear mode in generalized theories of gravity in
d+1 dimensions. It is shown that the shear diffusion coefficients of the
near-horizon and  boundary hydrodynamics are equal and can be
expressed in a form that is purely local to the horizon. We find that the
Einstein-theory relation between
the shear diffusion coefficient and the shear viscosity to
entropy ratio is modified for generalized gravity theories: Both
can be explicitly written as
the ratio of a pair of polarization-specific gravitational
couplings but implicate differently polarized gravitons.
Our analysis is restricted to the shear-mode
fluctuations for simplicity and clarity; however, our methods can be
applied to the hydrodynamics of all gravitational and matter
fluctuation modes.
}

\maketitle



\label{intro}
\section*{}

The long-wavelength fluctuations of the near-horizon metric of a
black brane in Anti-de Sitter (AdS) space  and the long-wavelength
fluctuations of the metric near the AdS boundary  can  each be
described by a  translation-invariant (effective) thermal field
theory.
The equations of motion of either of these theories are  hydrodynamic
equations \cite{hydroI}.
The relevant parameters --- such as the
temperature, shear viscosity, diffusion coefficient
and entropy density --- are
 intrinsic properties of the horizon and, as
such, should be defined strictly in terms of the near-horizon metric.
In spite of this apparent locality, considerable
evidence has accumulated suggesting that these
same horizon-specified quantities should be used for the boundary
theory.
Indeed, there are some  concise expositions on this very point
\cite{hydroI,alex-liu,hydroIII,starinets-M}; with these
having been able to establish  the boundary field theory
as describing a viscous fluid with precisely these hydrodynamic
parameters.

The AdS boundary hydrodynamics can be related via the gauge--gravity
duality to the hydrodynamics of strongly coupled gauge theories
\cite{PSS1,hydroIII}. The latter provides an interesting theoretical
framework for studying relativistic hydrodynamics and may explain the
experimental results of
heavy-ion collisions, non-relativistic systems exhibiting
superfluidity, {\it etc}. \cite{ion}. The boundary hydrodynamics has
been
most  extensively studied using Einstein's theory of gravity, which
corresponds to infinitely strong 't Hooft coupling on the gauge-theory side.
As for applying the results to real physical systems that can be
described by gauge theories, one really needs to know the outcomes for finite
values of 't Hooft  coupling. Then, since
the strong-coupling expansion on the gauge-theory
side corresponds to
the derivative expansion on the gravity
side, such an extension actually requires knowledge of the results for
generalized theories of gravity. More specifically,
one would  first have to calculate the higher-derivative
corrections to Einstein's gravity in string theory and then
calculate the hydrodynamics of the corrected theory.

Here, we will provide a general prescription on how to perform the
second
stage for  arbitrarily general gravitational corrections.  First,
we will establish a more direct connection between
the near-horizon hydrodynamics and that of the
AdS boundary. This task will be accomplished
by showing that the thermal field-theoretic formalism developed for
the
boundary hydrodynamics can be directly translated into a
calculation that is completely local to the horizon.
Importantly, absolutely no information about
the hydrodynamics is lost in this localization process.
To extract the hydrodynamical parameters --- such
as the shear diffusion coefficient --- at the AdS boundary,
one solves the gravitational
perturbation equations with incoming boundary conditions on the
horizon and Dirichlet conditions on the outer boundary. However,
it will be clear that one can similarly place the Dirichlet boundary
conditions on radial shells anywhere in between the horizon and the
boundary, including on the stretched horizon. After accounting for
the effect of the gravitational redshift, one can see that the shear
diffusion coefficients on all the shells are equal.

Next, we will go on to demonstrate how the aforementioned formalism
can be extended to any generalized (or Einstein-corrected) theory
of gravity. The hydrodynamic parameters of interest can  be
readily identified in terms of different components of a (generally)
polarization-dependent gravitational coupling
$\kappa_{\mu\nu}$.  Recently, we have proposed a method for
calculating these
polarization-dependent couplings that is completely local
to the horizon \cite{us}.
This followed from an examination of the two-derivative, second-order
expansion of the
action for the gravitons ($h_{\mu\nu}$), as described in \cite{BGH}.
In particular, the gravitational coupling can be identified
on the basis that  the replacement
$h_{\mu\nu}\rightarrow \kappa_{\mu\nu}h_{\mu\nu}$ leads to a
canonical kinetic term for
the $\mu\nu$-polarized graviton.
(This identification will be  particularly relevant to the latter
stages of our analysis.)
It follows that any theory which is sensitive to the different
polarizations can be
expected --- for instance ---  to alter the
shear viscosity to entropy density ratio
 $\eta/s$ from its ``standard" (Einstein)
value of $1/4\pi$. This claim was put into quantitative terms with
the
following proposal \cite{us}:
$
\frac{\eta}{s}=\frac{1}{4\pi}\ \frac{ \left(\kappa_{rt}\right)^2}
{\left(\kappa_{xy}\right)^2}\;,
$
where the subscripts on the gravitational couplings denote
the implicated polarizations (to be defined
more rigorously below).
For Einstein's theory
or any theory  related to Einstein's by a field redefinition,
$\kappa^2_{\mu\nu}=\kappa^2_E={\rm constant}$ (to be precise, one half
of the $d+1$-dimensional Newton's constant).

Our particular objective  will be to determine the shear diffusion
coefficient $D$, for a generic gravity theory; this, by cognizance of it's
relation to  the pole structure of an appropriate thermal Green's
function.
It will be  shown that, in general, the product $DT$ differs from
the ratio $\eta/s$; in  conflict with the prediction of Einstein's
theory.
But this is {\em not} at all in conflict with the basic principles at
work.
As  already stressed, each of the hydrodynamic parameters
in play is sensitive to different polarizations and, therefore,
will react differently to a non-trivial deviation
from Einstein's theory. What is most significant is
the direct correlation between any of these parameters and
the gravitational coupling for  a very specific
class of gravitons.

We will, for illustrative purposes, be somewhat specific about
the choice of spacetime geometry and class of perturbations; however,
our procedure (and outcomes) can readily be repeated
for many other interesting cases.

Let us now proceed with the proposed analysis and
consider a black $p$--brane in a $d+1$-dimensional
(asymptotically) AdS
spacetime. (Note that $d = p+2 \ge 5$.) The associated  metric is
usually depicted in the
Schwarzschild-like form
$
ds^2=-\frac{r^2}{L^2}f(r)dt^2
+\frac{L^2}{r^2}\frac{dr^2}{f(r)}+\frac{r^2}{L^2}\left(\sum_i^p
dx_i^2\right)\;,
\label{1}
$
where $L$ is the AdS radius of curvature, $f(r)=1-\left( r_h/r
\right)^{p+1}$ and
$r_h$ locates the black brane horizon. A simple change of
coordinates,
$u=r_h^2/r^2$, gives us another useful form
\begin{equation}
ds^2=-\frac{r_h^2}{L^2 u}f(u)dt^2
+\frac{L^2}{4u^2}\frac{du^2}{f(u)}+\frac{r_h^2}{L^2u}\left(\sum_i^p
dx_i^2\right)
\;,
\label{2}
\end{equation}
where  $f(u)=1-u^{\frac{p+1}{2}}$. The horizon and (outer) boundary
are now
located at
$u=1$ and $u=0$ respectively, and $T=(p+1)r_h/4\pi L^2$ can be
identified
(in any coordinate system) as the Hawking temperature.

To study brane hydrodynamics, one expands the metric, $g_{\mu\nu}
\rightarrow {\overline g}_{\mu\nu} +h_{\mu\nu}$.
In accordance with the standard conventions, the coordinate $z$
is singled out as the direction of propagation of the graviton
on the brane and  one of the remaining transverse  directions is
denoted by $x$ ---
any of which are interchangeable  by virtue of the spatial isotropy
of the brane. Obviously, any of the $x$ coordinates could have been
picked instead of $z$ for the same reason.
Given  these conventions, $h_{\mu\nu}\sim {\rm exp}[-i\Omega t +i Q
z]$ (otherwise depending only on $u$),  where
$(\Omega,0,...,0,Q)$ is the $p+1$--momentum of the graviton.

Under a suitable choice of gauge (namely, the radial gauge $h_
{u\alpha}=0$ for any $\alpha$),
it has been shown that the non-vanishing fluctuations separate into
three decoupled classes;
with these being commonly classified as the scalar, shear and sound modes
\cite{PSS2}.
The latter two are of particular interest, as the diffusion
coefficient for the shear viscosity can be
directly extracted  from the pole structure of the associated
correlator. Our current
attention will be directed towards the shear channel, as an analogue
calculation for the sound modes will yield the same basic outcome
(albeit
with some additional information)
but with significantly  more technical clutter. We will, however,
briefly
discuss the sound channel near the end of the article.

To determine the thermal correlator in question, one can proceed
exactly as
in \cite{dam,pavel,mas}, where much more elaborate discussions can be
found.
It is first necessary to identify a gauge-invariant combination
of the shear-mode fluctuations $H_{tx}=(-1/g_{tt}) h_{tx}$ and $H_
{zx}=(1/g_{zz})h_{zx}$:
\begin{equation}
Z=q H_{tx}+\omega H_{zx}\;.
\label{3}
\end{equation}
Here, $\omega=\Omega/2\pi T$ and $q=Q/2\pi T$ represent
a dimensionless frequency and wavenumber respectively. Importantly,
either of these parameters is vanishing (although not necessarily at
the same rate) in the so-called hydrodynamic limit.

Restricting to the radial gauge and expanding out the Einstein field
equations to the linear order of $Z$,
one then  schematically obtains
$
Z^{\prime\prime} + A(\omega,q^2 f,u)Z^{\prime}+B(\omega,q^2 f,u)Z=0
\;,
$
where a prime is a derivative with respect to the radial coordinate
$u$.
The coefficients $A$ and $B$ can be found in, for instance,
eq.~(3.14) of
\cite{mas} for general $p$ (although with conventions differing from
ours) and
eq.~(4.26) of \cite{pavel} for the $p=3$ case.
What is important, for our purposes, is  not necessarily the
explicit structure of the coefficients but that $q$ appears uniquely
in the combination
$q^2 f(u)$. (Recall that $f(u)$ appears in the metric (\ref{2}).)

One is instructed to solve this differential  equation subject to a
specific pair of
boundary conditions. Firstly, the solution is constrained
to be an incoming plane wave at the horizon $u=1$.
To the linear order
of a perturbative expansion in $\omega$ and $q$, this  condition
imposes
a solution of the form \cite{pavel,mas}
\begin{equation}
Z=Cf(u)^{-\frac{i\omega}{2}}\left[1+\frac{iq^2}{2\omega}f(u)
+{\cal O}\left(\frac{q^4f^2}{\omega^2}\right)\right]\;.
\label{5}
\end{equation}
Here, $C$ is an integration constant that will  be fixed by imposing
the appropriate normalization condition.

Secondly,  there is the  so-called Dirichlet boundary condition,
which has yet to be enforced.  One can impose this condition
at any point $0 \le u^* < 1$; although it has become standard
procedure
to choose
$u^*=0$ and, thus, single out the boundary of AdS as a preferred
place.
However, this choice is not imperative.  What the condition
does  necessitate is
that $Z(u)$ ({\it prior}
to its normalization) is vanishing as $u\rightarrow u^*$, which in turn
imposes that $\omega=-iq^2 f(u^*)/2$ --- {\it cf}, eq.~(\ref{5}). As
an immediate consequence, we see that (in spite of first appearances)
$\omega$ and $q^2 f(u^*)$ are
of the same order in the hydrodynamic limit; that is, $\omega^2 <<
q^2 f
(u^*)$.
(This last observation is very important to the discussion that
follows.)
One then further requires that $Z$  be normalized
to unity at $u^*$.
For the case in hand, this can be achieved by the choice
\begin{equation}
C^{-1}=  \left[1+\frac{iq^2}{2\omega}f(u^*)\right]\;.
\end{equation}

At this point, let us make it clear that the hydrodynamic
or zero-frequency limit is supposed to be put into effect.
Once this limit
has been satisfied,
it  becomes a straightforward
exercise to show that both $Z$ and its  correlator (see directly
 below) are
radial invariants. (For a clear demonstration of this
invariance, see \cite{liu}.)

To learn about the  two-point functions of the stress tensor, it is
sufficient to study
the correlator $G_{ZZ}$ of the gauge-invariant variable $Z$. This
correlator is  directly
extractable from the boundary gravitational action. To elucidate,
one identifies  the
boundary residue of
the canonical term in the bulk action, $ZZ^{\prime}$,
as the correlator up to an inconsequential
numerical factor. Following
along these lines, one is able to deduce that
$
G_{ZZ}=-\lim\limits_{u\rightarrow 0} K \frac{f(u)}{u[\omega^2-f(u)
q^2]}Z(u)Z^{\prime}(u)\;,
$
where $K$ is a dimensional constant that depends only on
the metric length scales ($r_h$ and $L$)
and the gravitational coupling.
Then, taking  $u$ and $u^*$ to zero at the end
of the calculation (and employing the  Dirichlet condition to
simplify the differentiation), one finds that
$
G_{ZZ}=K \frac{iq^2}{\left[\omega^2-q^2\right]\left[\omega
+iq^2/2 \right]}\;
$
or, by invoking $\omega^2\ll q^2$,
$
G_{ZZ}=K \frac{1}{\left[i\omega-\frac{q^2}{2}\right]}\;.
$

One  can immediately notice the pole in correlator,
and, hence, the associated divergence. This singularity is not
undesirable;
rather, it can be viewed as a well-motivated
expectation from the quasinormal-mode perspective of brane
hydrodynamics \cite{pavel}.
Moreover, by using the standard hydrodynamic dispersion equation
$
\omega=-iDk^2+{\cal O}(k^4)\;
$
and recalling the previous scaling relations ($\omega=\Omega/2\pi T$
and
$q=Q/2\pi T$), we can
determine the diffusion coefficient as $D=1/4\pi T$.
The diffusion coefficient allows us, in turn, to fix the shear
viscosity  to entropy density ratio: $\eta/s=DT=1/4\pi$
 in agreement with the ``usual"
(but, as discussed earlier, not necessarily universal)
outcome for brane hydrodynamics.

Let us now restrict our attention to near-horizon physics.  After
all,
the stretched horizon --- defined here as the region $1-u^*\ll 1$
---  is a
perfectly legitimate choice.
So far, we have made the claim that the Green's function should
have the same functional form at both the stretched horizon and
boundary
(and all points in between), while acting as a thermal correlator
for a uniquely specified
hydrodynamic system. Let us  see how this actually plays out
by repeating the calculation for
$G_{ZZ}$ but, this time, taking the near-horizon limit; that is,
imposing  $u,\ u^*\rightarrow 1$ (at the very end of the calculation).
As before, all the unwarranted zeros in the calculation
nicely cancel, leaving
\begin{eqnarray}
G_{ZZ}&=& \lim_{u,u^*\rightarrow 1} K \frac{iq^2f(u)}{\left[\omega^2-
q^2f(u)\right]\left[\omega
+iq^2f(u^*)/2 \right]} \nonumber \\
&=&
\lim_{u,u^*\rightarrow 1} K \frac{1}{\left[i\omega
-\frac{q^2f(u^*)}{2} \right]}\;,
\label{11}
\end{eqnarray}
where the second line follows from the hydrodynamic limit.
Comparing the boundary result and (\ref{11}), one can immediately see
a difference;
namely, the would-be pole structure has changed by a factor of
$f(u^*\rightarrow 1)$.

We can address this new development as follows:
The frequency $\omega$ and wavenumber
$q$ are  coordinate-dependent constructs that will naturally
experience the effect of a radially dependent gravitational redshift.
So, how can we quantitatively  discern the
{\em relative} redshift given the sensitive (non-linear) relation
between $\omega$ and $q$?
The answer is remarkably simple: It is the structure of the
correlator
itself that tells
us exactly how to determine the relative redshift!
The  hydrodynamic limit is, in actuality,
an expansion in the ratio $q^2f(u)/\omega$ --- {\it cf}, eq.~(\ref{5}) ---
so that the limiting procedure will inevitably break down unless $q^2f(u^*)/\omega < 1$.
Let us also observe that $f(u)$ is monotonically increasing from its
horizon value of $f(1)=0$ to its boundary value of $f(0)=1$.
This means that, once imposed at the stretched horizon,  the hydrodynamic
expansion  at an arbitrarily larger radius  is only ensured to persist  if
$q^2/\omega$ scales as $1/f(u)$.

Taking $\omega$ to be  fixed and unaffected
by the redshift, so as not to disturb the incoming
(horizon) boundary condition --- which, when
properly enforced, fixes a precise form for the solution
at the surface of vanishing $f$ --- we then have that
$\omega(u)=\omega_{b}$
and $q^2(u)=q^2_b/f(u)$. (The subscript $b$ indicates
the outer boundary value of the redshifted quantities.)
Now, after applying the Dirichlet condition at the stretched
horizon
to the wavefunction (\ref{5}), one
can deduce a pole structure  of $\omega_{b}=-iq_{b}^2/2$ and,
accordingly, a diffusion coefficient of  $D=1/4\pi T$. In this way,
we have finally achieved full compliance between the
horizon and boundary calculations!

Moreover, it should be
evident that, from the current perspective, there
is really nothing special about either the outer
boundary or the black brane horizon. Which is to say,
the Dirichlet-imposing surface could have, just as
well, been placed at any point between the
stretched horizon and outer boundary without
jeopardizing the form of the pole structure
and, hence, the value of the diffusion coefficient.

Next, let us extend  considerations
from Einstein gravity to a general
theory of gravity. We will proceed
to show that, for a general
theory, the diffusion coefficient
is modified  in a very precise way.
This form will be verified by two independent
calculations; one of which is based on
extracting the (modified) pole of the
previously examined correlator and
a second which considers the
diffusion coefficient
as a proportionality constant in
a conservation equation for the
(dissipative) stress-energy tensor.

By a generalized gravity theory, we have in mind
a Lagrangian
$
\mathscr{L}=\frac{1}{32\pi\kappa_E^2}\left[R+
\lambda \mathscr{L}_C\right]\;
$
that allows for black brane solutions of the form (\ref{2}).
Here,  $\mathscr{L}_C$
represents  some
correction to Einstein's Lagrangian and
$\lambda$ is a constant ``tracking" parameter.
Formally speaking, the
correction
need not be perturbative for our framework to apply. However,
most (if not all) interesting cases in the literature
can be formulated as such. For further details and explanations,
the reader can consult \cite{us} (also, \cite{BGH}).

For  a generalized gravity theory,  the effective coupling can be
expressed as
$
\frac{\left(\kappa_E\right)^2}{\left(\kappa_{\mu\nu}\right)^2}=
1\mp\frac{\lambda}{2}
\left( \frac{\delta\mathscr{L}_C}{\delta R_{ab}^{~~cd}}\right)^{\!\!
(0)}
\hat\epsilon_{ab}\hat\epsilon^{cd}\;,$ where
$\{a,b,c,d\}\in\{\mu,\nu\}
$
and $\hat\epsilon_{ab}$ is the binormal vector with regard to the
specified
pair of polarization directions.
Any binormal is antisymmetric under the exchange of $a$ and $b$,
and normalized such that $\hat\epsilon_{ab}\hat\epsilon^{ab}=\mp 2$.
A $\mp$ sign
is only to be taken as negative when one of the directions
$(\mu,\nu)$
is timelike (this  convention ensures the positivity of the
coupling).
The superscript $(0)$ signifies that the calculation is always made
on solution {\em and} on the horizon.
Note that, at the order of the two-derivative expansion,
the generalized couplings can be treated as (polarization-dependent)
constants.

The basic premise now goes as follows:
If Einstein's gravity is ``non-trivially" modified (meaning that the
corrected theory
can {\em not} be  related to Einstein's theory by a  field
redefinition),
then the gravitational coupling is no longer as simple
as $\kappa^2_E$  and can be expected to depend on the
polarization of the gravitons being probed. For example, calculations
of the  black brane entropy are known to be sensitive to the $r-t$
(or $u-t$) polarized gravitons.
Meanwhile, for the other hydrodynamic parameters of interest, the
polarization directions
will depend on the particular channel being probed. More
specifically, the scalar channel
depends on  the $x-y$ (polarized) gravitons, the shear channel
depends on the $x-t$ and $x-z$
gravitons, and the sound channel depends on a particular combination
of $t-t$, $t-z$, $x-x$, $y-y$,
$z-z$ and $x-y$ fluctuations. (As before, $z$ represents the
direction of
propagation on the brane, while $x$ and $y$ are any other transverse
directions.)

We are now well equipped to discuss the diffusion coefficient for a
generalized gravity theory. Our first method is based on looking at
the pole structure of the correlator $G_{ZZ}$. One might well ask
as to
how  this pole would be modified
for a generalized theory. To answer this query, let us recall
the identification of the gravitational coupling,
$h_{\mu\nu}\rightarrow \kappa_{\mu\nu}h_{\mu\nu}$, as
discussed earlier in the article.
On this basis, it is quite natural to
modify eq.~(\ref{3}) for the gauge-invariant variable $Z$ as follows:
\begin{equation}
Z=q\kappa_{tx}H_{tx}+\omega\kappa_{zx}H_{zx}\;.
\end{equation}

One can now discover the correct scaling properties of
the hydrodynamical parameters in a way that resembles dimensional
analysis: {\em First}, redefine the  wavenumber and (in principle)
the
frequency with a scaling operation, {\em second},
reformulate the field equation and  then the
solution in terms of these rescaled parameters and, {\em third}, read
off the revised pole structure.
Before proceeding with the first step, let us
fix $\omega$ (as previously explained) to preserve the incoming boundary
condition at the horizon. Then,
by exploiting the freedom (at this level of analysis) to change
the normalization of $Z$, we obtain
$
Z= q\frac{\kappa_{tx}}{\kappa_{zx}}H_{tx}+\omega H_{zx}
= \widetilde{q}\ H_{tx}+\omega H_{zx}\;,
$ with ${\widetilde q}\equiv q\frac{\kappa_{tx}}{\kappa_{zx}} $.
Since the couplings can be treated as constants, the
solution (\ref{5}) remains basically unchanged and needs only to be
rewritten in terms of the rescaled parameter; that is,
$
Z\sim f(u)^{-\frac{i{\omega}}{2}}\left[1+\frac{i{\widetilde {q}}^2f
(u)}{2{\omega}}
\frac{\kappa^2_{zx}}{\kappa^2_{tx}}\right]\;.
$
Now, applying the Dirichlet boundary condition just
like before,
we can readily extract the  diffusion coefficient
\begin{equation}
D= \frac{\kappa^2_{zx}}{\kappa^2_{tx}}\frac{1}{4\pi T}\;
\label{15}
\end{equation}
for a generalized theory of gravity.

For a second method,
our result (\ref{15}) can be shown to  agree with the  expectations
of hydrodynamics as per the following argument:
In hydrodynamics, one can obtain $D$ by inspecting the $x$ component
of the
conservation equation $\partial_\mu T^{\mu x}=0$ for the dissipative
stress-energy tensor.
(The relevant stress tensor is that of the $p+1$-brane
theory. See \cite{hydroIII} for a topical discussion.)
Given the symmetries of the problem, this conservation equation
reduces to $\partial_t T_{tx}=\partial_z T_{zx}$.
The stress-energy tensor can then be
expressed in terms of derivatives of the
local fluid velocity $u_{\mu}$.
In linear hydrodynamics, $T_{zx}= - \eta \partial_z u_x$ whereby
$\eta$
is the same shear viscosity that appears in $T_{xy} =- \eta
\partial_y u_x$. To lowest order, $T_{tx}=-(\rho+p) u_x$ such that
$\rho$
is the energy density of the fluid and $p$ is its pressure.
Now, {\em equilibrium thermodynamics} would  imply
that $T_{tx}=-sT u_x\;$, leading to
$\partial_t u_x = \frac{\eta}{s T} \partial_z^2 u_x$  and then
$DT=\eta/s$. In general, however, this is not the case. Rather,
$T_{tx}= -(\rho+p+\delta) u_x \equiv \chi u_x\;$, with $\chi$
being the coefficient of
heat conductance. (The correction $\delta$ is a purely relativistic
effect that allows for an isothermal flow of heat in accelerated
matter
in the  direction opposite  to the acceleration
\cite{eckart,weinberg}.)  Combining these
equations, one then obtains
$\partial_t u_x = \frac{\eta}{\chi} \partial_z^2
u_x$.
Now, using the Kubo formula,
one would find  $\chi\sim 1/\kappa^2_{tx}$
in the same way that $\eta\sim 1/\kappa^2_{xy}$.
So that, appropriately scaling $\eta\to \eta (\kappa_{zx}/\kappa_E)^2$ and
$\chi\to \chi (\kappa_{tx}/\kappa_E)^2$, one
would make $D$ scale as $D \to
\left(\frac{\kappa_{zx}}{\kappa_{tx}}\right)^2 D $ --- exactly as in
eq.~(\ref{15}).

Let us briefly point out that analogous arguments
can similarly be applied to the sound channel;
from both  the thermal correlator and hydrodynamic perspectives.
A detailed account of the sound-mode calculations will be reported in a separate manuscript \cite{new}.

Finally, the informed reader might be concerned about an apparent
conflict
between our result and that of \cite{myers}. The authors of this
paper
studied a model with the (non-trivial) correction
$\lambda \mathscr{L}_C= \lambda R_{\mu\nu\sigma\rho}R^
{\mu\nu\sigma\rho}$
and found that $DT=\eta/s$ for
$d+1=5$ (or $p=3$). We will now show that this agreement is actually
a numerical coincidence specific to this particular dimensionality
but is not generally true.
As shown explicitly in our previous work \cite{us} (where we
carefully analyzed
the very same model),
$
\frac{1}{\left(\kappa_{rt}\right)^2}= \frac{1}{\left(\kappa_E\right)
^2}
\left[1+\frac{2\lambda}{L^2}d(d-3)\right]
$  and $\kappa_{xy}^2=\kappa_E^2\;$.
The same  basic procedure can also be applied to obtain
$
\frac{1}{\left(\kappa_{tx}\right)^2}= \frac{1}{\left(\kappa_E\right)
^2}
\left[1-\frac{2\lambda}{L^2}d\right]\;
$
and, of course,  $\kappa_{zx}^2=\kappa_E^2$.
(It should be noted that none of these outcomes depends
upon the particular choice of radial coordinate and, moreover,
$\kappa^2_{u\nu}=\kappa^2_{r\nu}$ for any $\nu$.)
It is now straightforward to compute and compare the ratios of
interest.
Namely, from $\frac{\eta}{s}=\frac{1}{4\pi}\ \frac{ \left(\kappa_{rt}
\right)^2}{\left(\kappa_{xy}\right)^2}$ and eq.~(\ref{15}), we
respectively obtain
$
\frac{\eta}{s}
= \frac{1}{4\pi}\left[1-\frac{2\lambda}{L^2}d(d-3)\right] \;
$
and
$
DT = \frac{1}{4\pi}
\left[1-\frac{2\lambda}{L^2}d\right] \;.
$
Clearly, the agreement that was found between these two ratios
is specific to the dimensionality
$d+1=5$. Note that, under even more general circumstances, we could not
expect
such agreement for any dimensionality.

\section*{}
{\bf Acknowledgments:}
The research of RB was supported by The Israel Science Foundation
grant no 470/06.
The research of AJMM is supported by the University of Seoul.

\end{document}